# Background-free Tracking of Ultrafast Hole and Electron Dynamics with XUV Transient Grating Spectroscopy

*Vincent Eggers[1,†,*], Rafael Quintero-Bermudez[1,2,‡], Kevin Gulu Xiong[1], Stephen R. Leone[1,2,3]*

[1]Department of Chemistry, University of California, Berkeley, California 94720, USA

[2]Chemical Sciences Division, Lawrence Berkeley National Laboratory, Berkeley, California 94720, USA

[3]Department of Physics, University of California, Berkeley, California 94720, USA

[†]Present Address: Department of Physics and Regensburg Center for Ultrafast Nanoscopy (RUN), University of Regensburg, 93040 Regensburg, Germany

[‡]Present Address: Apple, Inc., One Apple Park Way, Cupertino, CA 95014, USA

*vincent.eggers@ur.de

**ABSTRACT**

Extreme ultraviolet (XUV) transient absorption (TA) and transient reflectivity (TR) spectroscopies enable element-specific insights into attosecond-timescale processes in solids. XUV transient grating spectroscopy (TGS) is an emerging tool that combines the advantages of both absorption and reflectivity while offering intrinsically background-free detection. Here, we implement XUV-TGS by generating a transient grating in germanium solid using two few-cycle near-infrared pulses and probing it with an attosecond XUV pulse, produced via tabletop high-harmonic generation. The spectrally resolved, diffracted XUV pulses directly visualize the separate ultrashort decay times of both photoexcited electrons and holes, without the need for iterative deconvolution. By combining XUV-TA and -TG spectroscopy, we extract the evolution of the complex refractive index, $\tilde{n}$, without the need for Kramers–Kronig reconstruction, as required in XUV-TR, allowing us to extract the roots of the induced optical response. We find reflectivity changes of up to 34% via the real part of $\tilde{n}$, whereas changes in the imaginary part only result in a variation in reflectivity of around 0.5%.

**INTRODUCTION**

Understanding the ultrafast dynamics of photoexcited carriers in solids is essential for advancing technologies in optoelectronics, photovoltaics, and quantum information [1-3]. Processes such as carrier generation, scattering, and recombination occur on femtosecond to attosecond timescales and are often element-specific, necessitating techniques that combine high temporal resolution with chemical sensitivity [4,5]. XUV transient absorption and transient reflectivity spectroscopies have emerged as powerful tools for probing these dynamics down to sub-femtosecond timescales [6]. These methods leverage attosecond XUV pulses generated via the process of high harmonic generation (HHG) to access core-level transitions, offering direct insight into the evolution of electronic states [7-10]. However, both TA and TR rely on intensity changes of the transmitted or reflected probe and thus suffer from inherent background signals, small changes in differential measurements, and complex data processing, such as iterative decomposition [11] or Kramers–Kronig reconstruction [12], rendering quantitative extraction of the material responses challenging and often reliant on prior knowledge of the involved processes.

TGS addresses these limitations by combining the strengths of absorption and reflectivity techniques while offering intrinsically background-free detection by wave vector (phase) matching. By interfering two noncollinear excitation pulses with the same relative phase, a transient grating (TG) of carrier density is generated in the sample. The diffraction of a time-delayed probe from this grating encodes the transient modulation of the optical properties, while naturally rejecting any background light field that is not propagating into the diffracted direction [13]. Adding to efforts establishing XUV pump–XUV probe TGS at free-electron laser facilities [14-15], recent developments have established XUV-TGS as a tabletop technique combining the spatial selectivity and background-free detection of four-wave mixing with the element specificity and high temporal resolution of attosecond XUV probing [16]. This renders the method a uniquely suited tool for resolving ultrafast dynamics in solids.

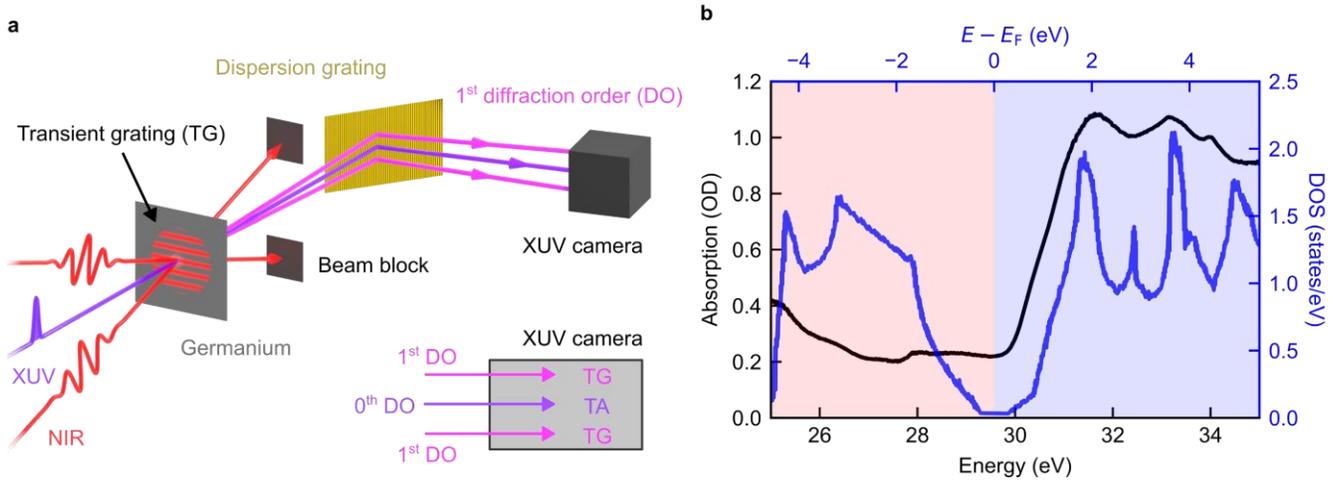

**Figure 1.** Extreme UV transient grating spectroscopy in germanium. (a) Schematic of the experimental setup. (b) XUV linear absorption spectrum (black line) and projected density of states (DOS, blue line) in the vicinity of the $M_{4,5}$ edge of germanium (~ 30 eV). Around this photon energy, the XUV light probes both the valence bands (red shaded area) and the conduction bands (blue shaded area) by promoting electrons from the 3d-core states to unoccupied states.

**EXPERIMENTAL METHOD**

In this work, we apply XUV-TGS to directly track the ultrafast electron and hole dynamics in germanium (Ge). Germanium is a prototypical semiconductor with many technical applications ranging from electronics to sensors and solar cells [17-19]. It has been widely studied with both TA- and TR-XUV spectroscopy to follow carrier relaxation and injection dynamics on ultrafast timescales [10-12]. Here, we employ two few-cycle NIR pulses to generate a spatially modulated excitation and, by probing the resulting transient diffraction of an attosecond XUV pulse, we spatially isolate the carrier-induced modulation of the dielectric function without disturbance from the unperturbed probe field. This approach directly links the diffracted XUV amplitude to the complex optical response of the material and enables direct visualization of electron and hole dynamics without the need for iterative deconvolution methods. In addition, combining XUV-TGS with conventional XUV-TA allows us to infer changes in the complex refractive index, without the need for Kramers–Kronig reconstruction.

The experimental geometry is illustrated in Fig. 1a. A detailed description of the setup can be found in [16]. A portion of the output of a Ti:sapphire amplifier (7 mJ maximum pulse energy, 35 fs pulse duration) is focused into a hollow core fiber to generate a supercontinuum spanning from 500 to 1000 nm, which is compressed to a pulse duration of 6 fs (Figure S1). A part of this pulse is branched off and focused into a Kr gas cell to generate a 25−45 eV attosecond XUV pulse (Figure S2), which is focused onto the sample as a probe beam. The remaining NIR energy is led through a 50:50 beamsplitter and then used to generate a grating on the sample. The two NIR beams are recombined with the time-delayed XUV beam and focused non-collinearly onto the sample with an angle of 36 mrad between the XUV and each NIR beam. The resulting NIR-NIR angle of 72 mrad yields a grating period of 11 μm (Figure S3) and the pump fluence on the sample is 26.5 mJ/cm$^2$ with a peak intensity of 4.1 TW/cm$^2$. Both the transmitted and refracted XUV beams are spectrally dispersed by a grating and guided to an XUV camera, where due to their spatial separation, both TG and TA can be simultaneously recorded. The sample consists of self-grown 50 nm thick films of Ge evaporated on 30 nm thick Si$_3$N$_4$ membrane substrates. Figure 1b shows the static absorption spectrum of Ge near the M$_{4,5}$ edge. Here, the XUV photons promote electrons from the 3d-core states into unoccupied states in the valence or conduction bands [11,20]. In the presence of the pump grating field, the NIR pulses can additionally promote electrons from the valence to the conduction band, resulting in a change of the material response in the spectral regions above and below the M$_{4,5}$ edge, corresponding to the conduction band and valence bands (projected DOS [21], Fig. 1b), respectively.

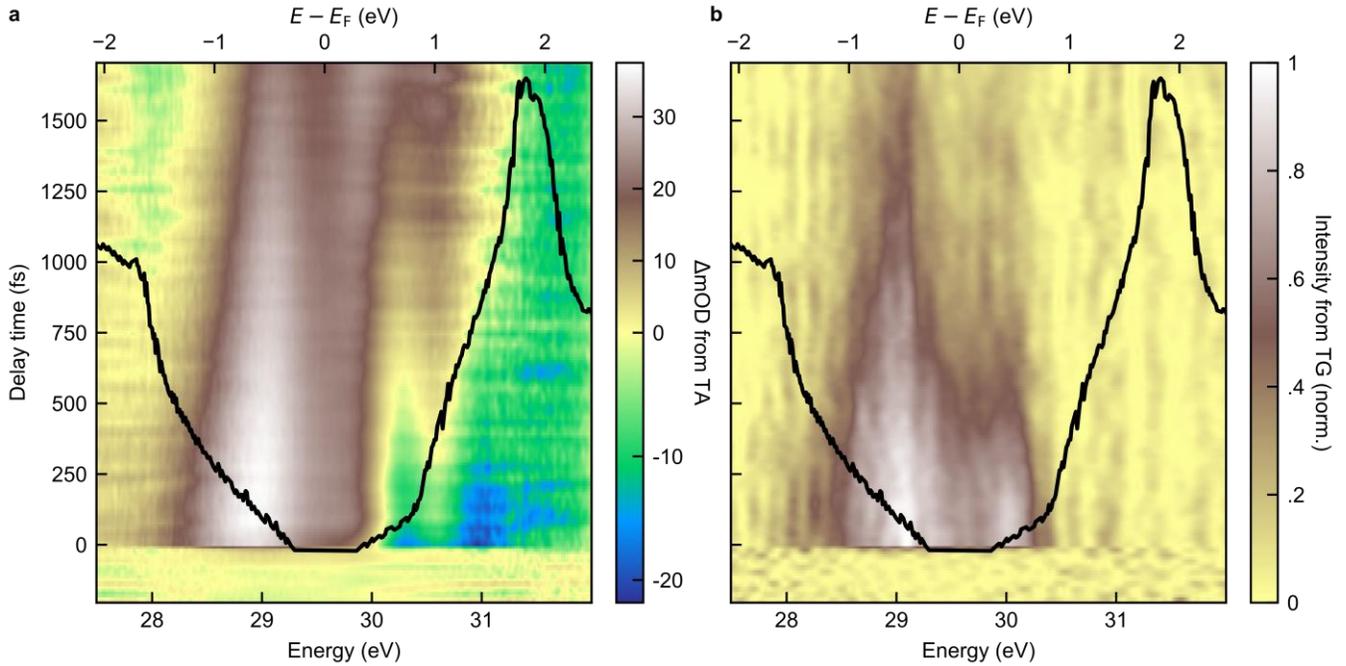

**Figure 2.** Time-resolved transient absorption and transient grating data. (a) The pump-induced changes to the transient absorption spectrum reveal positive and negative components spanning the spectral regions connected to both electrons and holes. The different features overlap partially, and they decay over a time span of a few ps. The black line shows the projected density of states as a reference. (b) The transient grating signal, naturally, has only positive components, resulting in a less complex image. It shows two main components with different decay rates, one in the spectral region associated with holes in the valence band, around 29 eV, and the other in the spectral region that relates to electrons in the conduction band, around 30 eV.

## RESULTS AND DISCUSSION

The pump-induced changes to the absorption are displayed in Fig. 2a. Transient absorption spectra for different time delays, $\tau$, show features with positive and negative signs in the observed spectral region, which start to decay over the observed time span from –200 fs to 1700 fs. Unfortunately, due to the many overlapping features, the high complexity of the data prevents immediate interpretation of the raw data map. In a previous TA work, the main contributors to the signal were identified by an iterative decomposition procedure [11]. The signal was largely reconstructed by considering state blocking, a broadening of the excited state, and a redshift of the ground state. Finally, accounting for the spin-orbit separation of the 3d-core states in the state blocking contribution, an iterative computational procedure was able to reveal the electron and hole dynamics. The main problem for that analysis is that the negative signal resulting from state blocking in the conduction band is superimposed with the signal corresponding to the redshift of the ground state, which increases absorption in the same spectral region. This precludes a direct extraction of the amplitudes and exact decay rates of the features connected to the electron dynamics from the raw data.

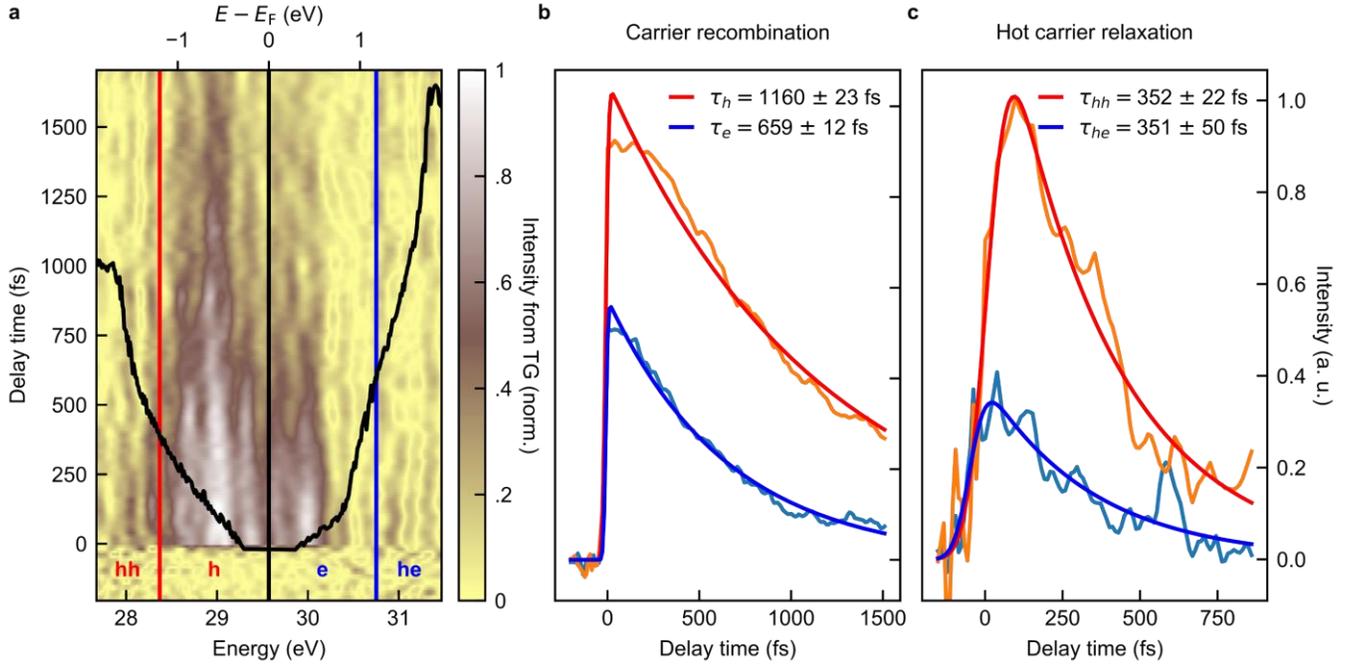

**Figure 3.** Electron and hole dynamics revealed by transient grating. (a) Modified transient grating signal accounting for spin-orbit separation. The map is divided into different spectral regions, which correspond to hot holes (hh) and hot electrons (he) and thermalized holes (h) and thermalized electrons (e). The black line shows the projected density of states as a reference. (b) Carrier recombination dynamics revealed by fitting the decay of the integrated signal in the electron and hole regions. (c) Hot carrier relaxation characterized by fitting the decay of the integrated signal in the hot electron and hole regions.

In contrast, the signal obtained by the transient grating (Fig. 2b) exhibits reduced complexity. Both in the region that can be attributed to holes, as well as in the region attributed to electrons, spectrally broad features appear that decay over a similar timespan as in the TA map. Since there is no background to the signal, there are only positive contributions in this case, making it much easier to directly compare features from different energy regions to each other. A feature associated with the redshift of the ground state is missing or weak. This reveals a difference in the decay rate of the signal from the electron and hole regions, which cannot be identified in the TA image.

After removing spectral features associated with spin–orbit separation using a Fourier filter applied to the data [22,11], a quantitative analysis of the electron and hole dynamics from the transient grating signal (Fig. 3a) is obtained. To differentiate between hot carrier and thermalized carrier contributions, the spectrum is separated into different regions. In line with previous work [11], the signal between $E - E_F = 0$ eV and $E - E_F = +(-)$ 1.2 eV is associated with thermalized electrons (holes) (Fig. 3b) and the region between $E - E_F = +(-)$ 1.2 eV and $E - E_F = +(-)$ 1.9 eV with hot electrons (holes) (Fig. 3c). Fitting the decay of the integrated signal in the respective regions reveals a carrier recombination time of $1160 \pm 23$ fs for the holes and a recombination time of $659 \pm 12$ fs for the electrons. The recombination time for holes is in good agreement with previous studies [11,12]. The observed difference in the recombination time of electrons and holes has previously been attributed to the different orbital characters of the conduction band valleys of Ge [12]. The energetically lowest lying L valley mainly has 4s character, rendering it invisible when probing with the XUV from the 3d-core states. This process should generally lead to an apparently faster decay of the electron signal, as electrons not only recombine with holes in the valence band but also scatter into the L valley. Further analysis yields a relaxation time of $352 \pm 22$ fs for the hot holes and $351 \pm 50$ fs for the hot electrons, which are also in good agreement with previous results [11,12].

The simultaneous collection of TA and TG data enables the analysis of the transient change of the refractive index without the need for Kramers–Kronig reconstruction [13]; the necessary procedure is detailed in the Supporting Information (SI). While heterodyne detection of the diffracted light is required to extract changes of the amplitude and phase of the complex refractive index, $\tilde{n}$, only the intensity of one of the diffraction orders was observed here. Still, with this, the absolute change of the real part of the refractive index, $n$, can be retrieved. In addition, changes of the imaginary part of the refractive index, $k$, can be calculated from the TA data. The extracted evolution of the refractive index is shown in Figure 4a and in Figure 4b, highlighting the dynamics of *n* and *k*, respectively. For *n*, the relative strength of the signal shifts towards the feature at 30 eV, for which previous studies observed a strong band shift [11,12]. In contrast, the map showing the changes to *k* is virtually identical to the TA data. We can further calculate the resulting changes to the reflectivity, *R*, from the pump-induced modulation of the complex refractive index, $\tilde{n}$, at a specified TR geometry, in this case with an incident angle chosen to be 66° versus normal (Fig. 4c,d). The result qualitatively reproduces comparable TR measurements [12]. It highlights the relative sensitivity of TR measurements to the real or imaginary part of $\tilde{n}$. We find that the variation in the real part of $\tilde{n}$ leads to relative changes of the reflectivity of up to 34% (Fig. 4c), while the variation of the imaginary part results in relative changes of the reflectivity of only up to 0.5% (Fig. 4d); this trend matches calculations for the sensitivity of TR to the different parts of the refractive index in Ref. 23.

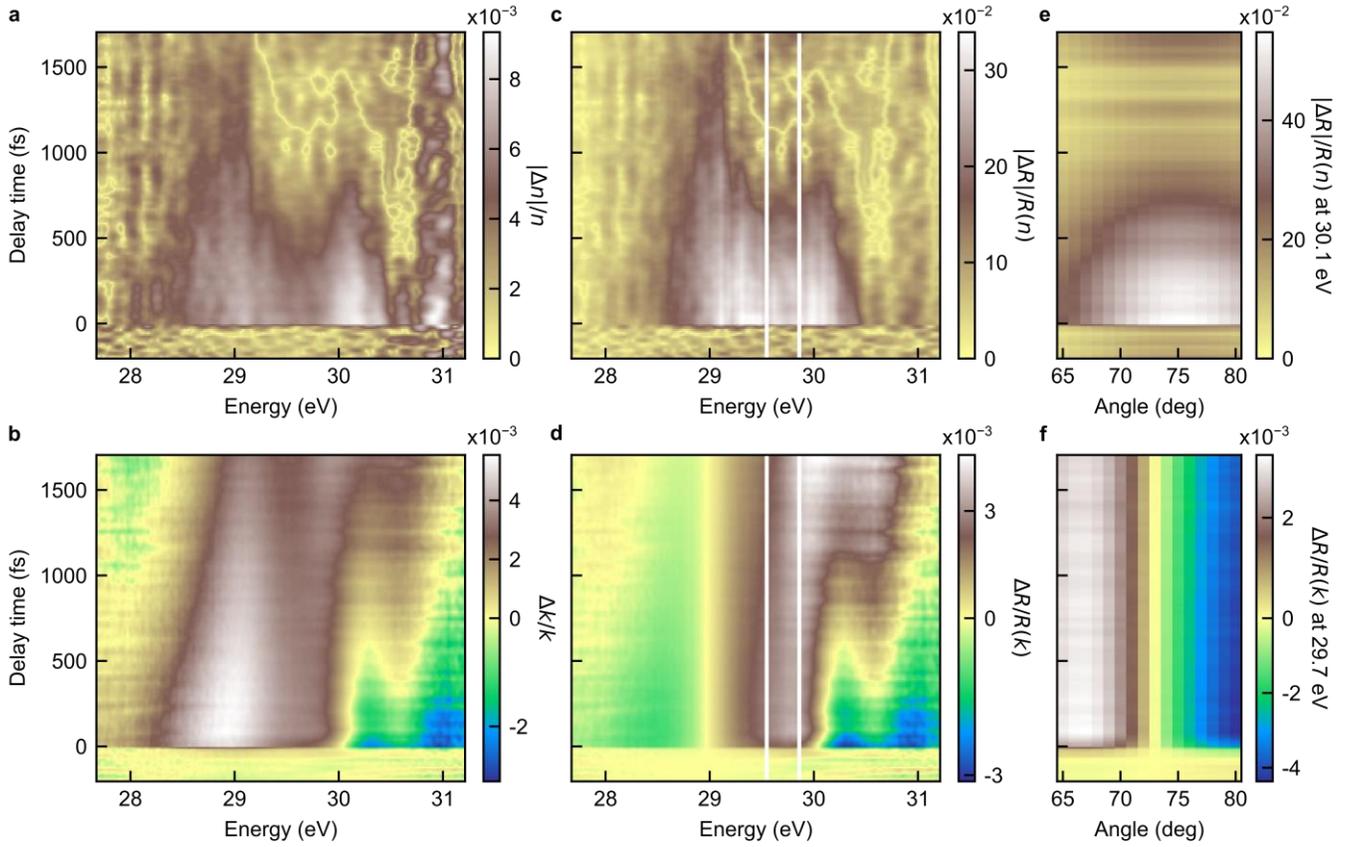

**Figure 4.** Pump-induced changes of the refractive index and reflectivity. (a) Absolute change of the real part of the refractive index, $n$, determined from calibrated TA and TG measurements. (b) Change of the imaginary part of the refractive index, $k$, determined from a calibrated TA measurement. (c) Absolute change of the reflectivity, $R$, reconstructed from the modulation of $n$ for an assumed incident angle of 66°. (d) Change of the reflectivity reconstructed from the modulation of $k$ for an incident angle of 66°. (e) Sensitivity of the reflectivity to changes in $n$ as a function of incident angle, characterized by integration of the region between the white lines in (c). (f) Sensitivity of the reflectivity to changes in $k$ as a function of incident angle, characterized by integration of the region between the white lines in (d).

By varying the angle in the calculations and focusing on characteristic features in the reflectivity map (solid white lines, Fig. 4c,d), a fundamental problem of TR measurements becomes apparent: The sensitivity of the reflectivity to changes in *n* reaches a maximum near the critical angle for total external reflection (Fig. 4e), while the sensitivity to changes in *k* approaches zero there (Fig. 4f). Together with the significantly lower sensitivity to changes in *k*, this poses a major challenge when attempting to extract the complete time-dependent complex dielectric function from TR data using Kramers–Kronig reconstruction, which additionally requires knowing both Δ*R* and *R* over a sufficiently large energy range [24]. The combination of TG and TA provides an excellent alternative to this scenario, as TG is sensitive to both *n* and *k*, and the reconstruction of the time-dependent complex refractive index does not rely on a priori knowledge. Furthermore, both TG and TA can be extracted with the same sample geometry, which greatly simplifies the experiments.

**CONCLUSION**

In summary, this work demonstrates the use of XUV-TGS to achieve background-free tracking of ultrafast electron and hole dynamics in germanium. By generating a transient grating with two few-cycle near-infrared pulses and probing with an attosecond XUV pulse, we directly visualize carrier relaxation and recombination without iterative deconvolution or Kramers–Kronig reconstruction. The combined use of XUV-TG and XUV-TA spectroscopy allows us to reconstruct the time-dependent complex refractive index and assess its impact on reflectivity, revealing fundamental limitations of conventional transient reflectivity techniques. Extreme UV transient grating spectroscopy emerges as a powerful tabletop technique for element-specific studies of ultrafast processes in solids. The ability to extract both real and imaginary components of the refractive index from XUV-TG and XUV-TA data opens new opportunities for quantitative investigations of carrier dynamics, the evolution of the dielectric function, and many other light–matter interactions on the femtosecond to attosecond timescale. Future work will extend this approach to more complex materials and explore heterodyne detection schemes [25] to fully retrieve amplitude and phase information of the complex refractive index.

## ASSOCIATED CONTENT

**Supplementary Information**

Pulse- and grating characterization; DFT calculation details; Conversion from the transient grating signal to the change of the refractive index. (PDF)

## AUTHOR INFORMATION

**Author Contributions**

S.R.L., R.Q.-B. and V.E. conceived the study. V.E., R.Q.-B., and K.G.X. carried out the experiment. R.Q.-B., V.E., and K.G.X. realized the experimental setup. V.E. analyzed the experimental data. V.E. prepared the samples. All authors analyzed and discussed the results. V.E. and S.R.L. wrote the paper with contributions from all authors.

**Funding Sources**

This work was supported by Air Force Office of Scientific Research (AFOSR) Grant FA9550-20-1-0334. K.G.X. and S.R.L. were partially supported by the Department of Energy, Office of Science, Basic Energy Science (BES) Program within the Materials Science and Engineering Division (contract DE-AC02-05CH11231), through the Fundamentals of Semiconductor Nanowires Program through the Lawrence Berkeley National Laboratory. S.R.L. also acknowledges sustained support by AFOSR Grant FA9550-24-1-0184 and an AFOSR DURIP grant FA9550-22-1-0451 for equipment. R.Q.B. acknowledges support from Natural Sciences and Engineering Research Council of Canada (NSERC) Postdoctoral Fellowship.

**Notes**

The authors declare no competing financial interest.

ACKNOWLEDGMENT

**Acknowledgment**

V. E. thanks Giacomo Inzani for fruitful discussions.

# Supplementary Information: Background-free Tracking of Ultrafast Hole and Electron Dynamics with XUV Transient Grating Spectroscopy

*Vincent Eggers[1,†,*], Rafael Quintero-Bermudez[1,2,‡], Kevin Gulu Xiong[1], Stephen R. Leone[1,2,3]*

[1]Department of Chemistry, University of California, Berkeley, California 94720, USA

[2]Chemical Sciences Division, Lawrence Berkeley National Laboratory, Berkeley, California 94720, USA

[3]Department of Physics, University of California, Berkeley, California 94720, USA

[†]Present Address: Department of Physics and Regensburg Center for Ultrafast Nanoscopy (RUN), University of Regensburg, 93040 Regensburg, Germany

[‡]Present Address: Apple, Inc., One Apple Park Way, Cupertino, CA 95014, USA

*vincent.eggers@ur.de

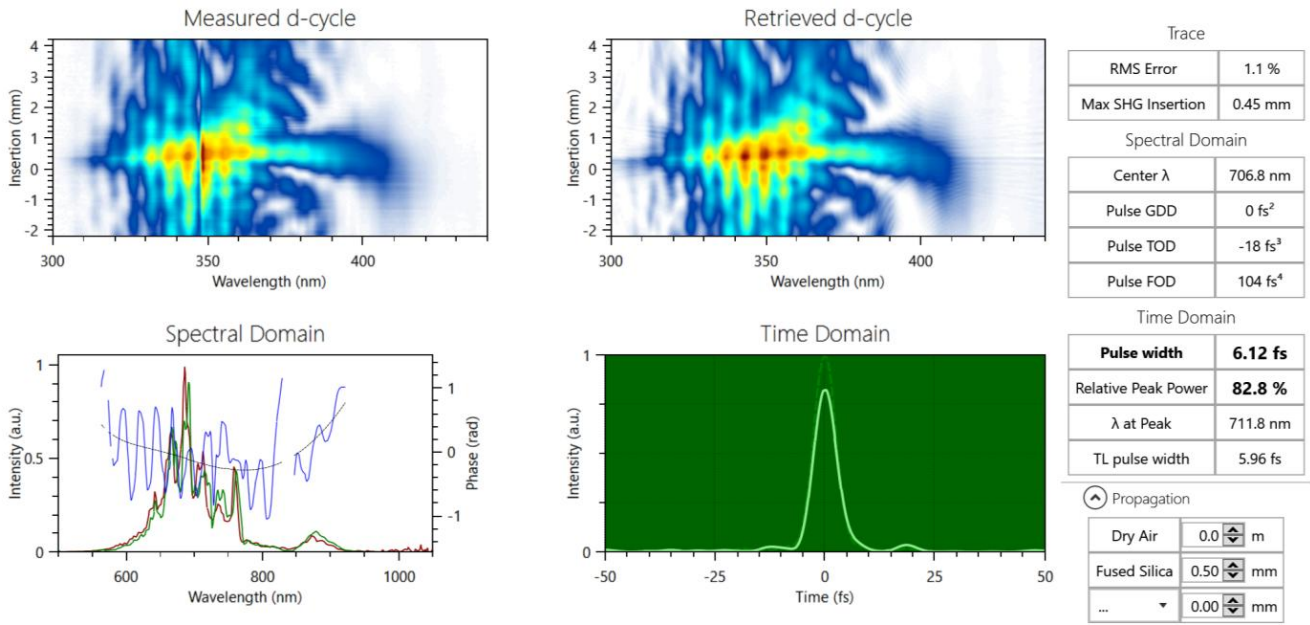

**Figure S1.** SHG dispersion scan measurement of the NIR pulses. Top left: Measured, calibrated d-cycle trace. Top right: Retrieved d-scan trace. Bottom left: Measured linear spectrum (red) and retrieved spectral phase (blue). Bottom right: Temporal intensity of measured pulse (light green) and transform-limited pulse (dashed green). GDD: Group delay dispersion; TOD: Third order dispersion; FOD: Fourth order dispersion; TL: Transform Limit; FWHM: Full width at half maximum.

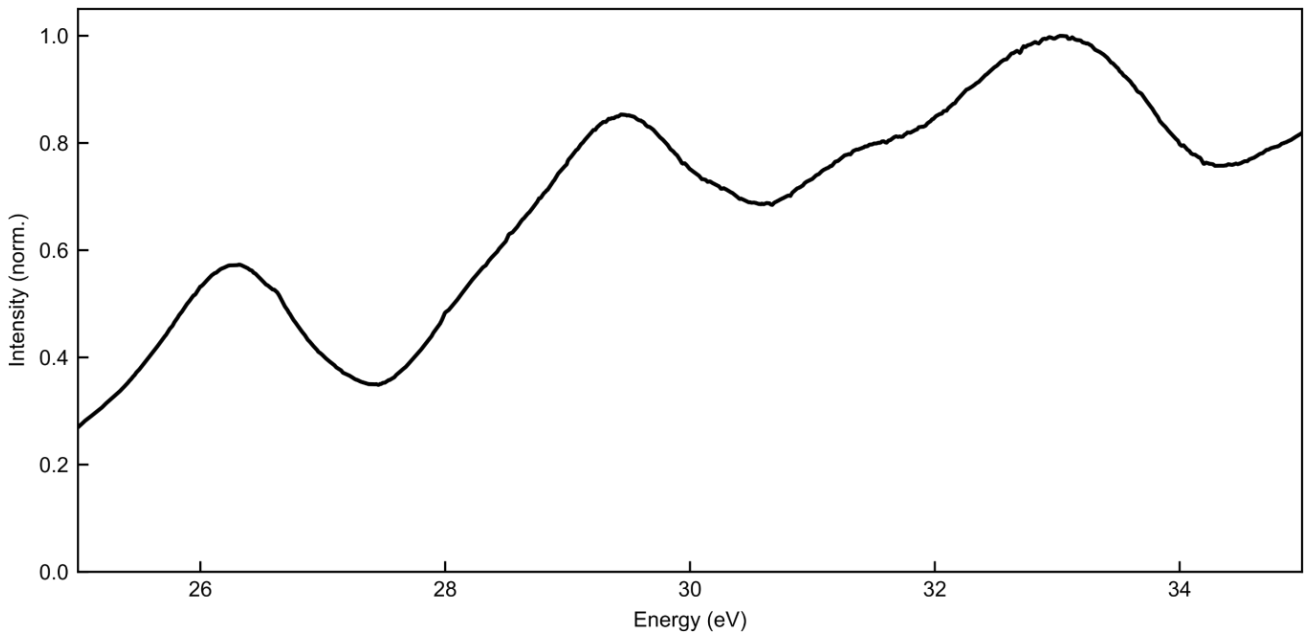

**Figure S2.** XUV probe spectrum. Spectrum of the HHG probe pulse before the sample.

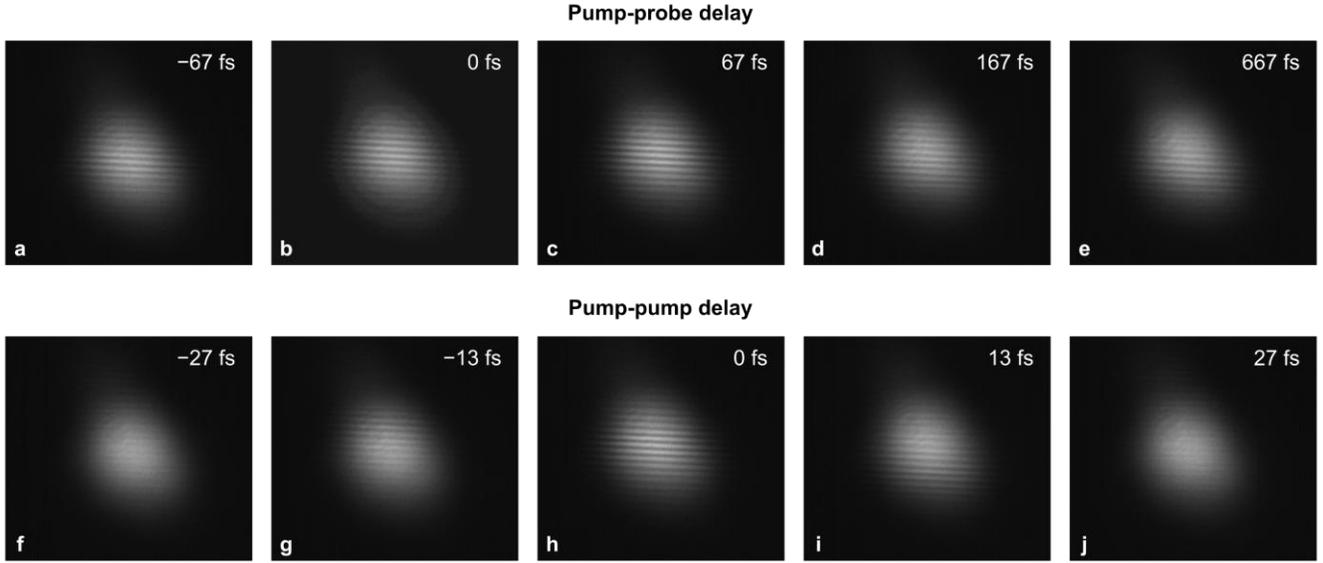

**Figure S3.** Stability of the transient grating. Transient grating recorded with a CCD camera at various pump-probe delays, (a-e) and at various pump-pump delays, (f-j). The gratings maintain good contrast when scanning pump-probe delay. However, when detuning the overlap of the two NIR pump pulses via the pump-pump delay, the grating quickly loses contrast and eventually vanishes.

**DFT calculations.**

The density of states (DOS) was adapted from the AFLOW materials database (ICSD #181071) [1], adjusting the energy gap such that characteristic features in the projected DOS align with previous experimental work [2,3].

**Calibration and conversion of the transient grating and transient absorption data.**

The transient grating (TG) efficiency, $\eta_{TG}$, connects the intensity of the first-order diffracted beam, $I_{TG}$, with the intensity of the incoming high harmonics (HH) radiation, $I_{HH}$. It is defined by [4]

$$\eta_{TG} = \frac{I_{TG}}{I_{HH}} = \left(\frac{\Delta\alpha d}{4}\right)^2 + \left(\frac{\pi \Delta n d}{\lambda}\right)^2,$$

with $\Delta\alpha$ being the change in the absorbance, $\Delta n$ the change in refractive index, $d$ the sample thickness, $\lambda$ the probe beam wavelength, and $\alpha$ the sample attenuation coefficient. Through a transient absorption (TA) measurement, we can extract the pump-induced change in absorbance, $\Delta\alpha$, by measuring the transmitted intensity under the influence of the pump beam, $I_{TA,\text{on}}$, and without the pump beam, $I_{TA,\text{off}}$, via

$$\Delta\alpha d = -\ln\frac{I_{TA,\text{on}}}{I_{TA,\text{off}}}.$$

With a TG measurement, we measure the transmitted TG intensity, $I_{TG,\text{trans}} = I_{TG} \cdot e^{-\alpha d}$. Taking a TA and TG measurement and measuring the HH intensity, therefore, allows us to calculate the absolute pump-induced change in the real part of the complex refractive index, $\tilde{n} = n + ik = \sqrt{\varepsilon}$, with

$$|\Delta n| = \sqrt{\frac{I_{TG,\text{trans}}}{I_{HH}} \cdot e^{\alpha d} - \left(\frac{\Delta\alpha d}{4}\right)^2}\,\frac{\lambda}{\pi d},$$

and the pump-induced change in of the imaginary part of the complex refractive index with

$$\Delta k = \Delta\alpha \cdot \frac{\lambda}{4\pi}.$$

For successful conversion, it is crucial that the pump fluence for the TA and TG measurements is constant, and that the collection efficiency remains the same when recording $I_{TG,\text{trans}}$ and $I_{HH}$. With this, it is possible to calculate the resulting changes in the reflectivity, $R$, at a chosen angle of incidence, $\theta_i$, via [5]

$$\frac{\Delta R(n)}{R} = 2\frac{\Delta n}{R} \cdot \text{Re}\left\{r_s^* \cdot 2\tilde{n}\frac{\partial r_s}{\partial \varepsilon_1}\right\},$$

and

$$\frac{\Delta R(k)}{R} = 2\frac{\Delta k}{R} \cdot \text{Re}\left\{r_s^* \cdot \frac{2\tilde{n}}{-i}\frac{\partial r_s}{\partial \varepsilon_1}\right\},$$

with the Fresnel coefficient

$$r_s = \frac{\cos\theta_i - \sqrt{\tilde{n}^2 - \sin\theta_i^2}}{\cos\theta_i + \sqrt{\tilde{n}^2 - \sin\theta_i^2}},$$

and

$$\frac{\partial r_s}{\partial \varepsilon_1} = \frac{-\cos\theta_i}{\sqrt{\tilde{n}^2 - \sin\theta_i{}^2}} \cdot \frac{1}{\left[\cos\theta_i + \sqrt{\tilde{n}^2 - \sin\theta_i{}^2}\right]^2} \ ,$$

with $\varepsilon_1$ as the real part of $\varepsilon$. The data for the reflectivity and complex refractive index were obtained from Ref. 6.